\documentclass[structabstract]{aa}  
\usepackage{amsmath}
\usepackage{natbib}
\usepackage{graphicx}
\usepackage{txfonts}
\newcommand{\kms}{\hbox{${\rm km}\:{\rm s}^{-1}\;$}}
\newcommand{\kmso}{\hbox{${\rm km}\:{\rm s}^{-1}$}}
\newcommand{\teff}{$T_{\rm eff}\;$}  
\newcommand{\teffo}{$T_{\rm eff}$}  
  
\newcommand{\tefo}{$T_{\rm eff}^4$}  
\newcommand{\logg}{$\log{g}\;$}   	
\newcommand{\loggo}{$\log{g}$}  	

\bibpunct{(}{)}{;}{a}{}{,} 
\begin{document}
\title{A new implementation of the infrared flux method \\ using the
2MASS catalogue}
\titlerunning{IRFM using the 2MASS catalogue}
\authorrunning{Gonz\'alez Hern\'andez et al.}
%
%
\author{J.~I.~Gonz\'alez Hern\'andez\inst{1,*}, \and P.
Bonifacio\inst{1,2,*}}
\offprints{J.~I. Gonz\'alez Hern\'andez}
\institute{
GEPI, Observatoire de Paris, CNRS, Universit\'e Paris Diderot; Place
Jules Janssen 92190
Meudon, France \\
\email{[Jonay.Gonzalez-Hernandez];[Piercarlo.Bonifacio]@obspm.fr}
\and
Istituto Nazionale di Astrofisica - Osservatorio Astronomico di
Trieste, Via Tiepolo 11, I-34143  Trieste, Italy
\thanks{
Cosmological Impact of the First STars (CIFIST) Marie Curie Excellence
Team}} 

\date{Received ... August 2008; accepted ... January 2009}
 
\abstract
{The effective temperature scale of FGK stars, especially at the
lowest metallicities remains a major problem in the chemical abundance
analysis of metal-poor stars.} 
{We present a new implementation of the infrared flux method (IRFM)
using the 2MASS catalogue.}
{We computed the theoretical quantities in the 2MASS $JHK_{\rm s}$ 
filters by integrating theoretical fluxes computed from ATLAS
models, and compare them directly with the observed 2MASS 
$JHK_{\rm s}$ magnitudes.
This is the main difference between our implementation of the
IRFM and that of Ram{\'\i}rez \& Mel\'endez (2005, ApJ, 626, 446;
hereafter RM05), since to introduce new stars at the lowest
metallicities they transform the 2MASS $JHK_{\rm s}$ magnitudes into
the TCS photometric system. We merge in our sample 
the stars from Alonso et al. (1996, \aaps, 117, 227; hereafter AAM96;
1999, \aaps, 139, 335; hereafter AAM99), and other studies to
appropriately cover a wide range of metallicities, ending up with 555
dwarf and subgiant field stars and 264 giant field stars.
We derived a new bolometric flux calibration using the available
Johnson-Cousins $UBV(RI)_{\rm C}$ and the 2MASS $JHK_{\rm s}$
photometry. We also computed new \teff versus colour empirical
calibrations using our extended sample of stars.}    
{We derived effectives temperatures for almost all the stars in
the AAM96 and AAM99 samples and find that our scales of temperature
are hotter by $\sim64$\,K ($\sigma_T=104$\,K, $N=332$ dwarfs) and
$\sim54$\,K with a $\sigma_T=131$\,K  ($N=202$ giants). The same
comparison with the sample of RM05 for stars
with [Fe/H]~$<-2.5$ provides a difference of $\sim -87$\,K 
($\sigma_T=194$\,K, $N=12$ dwarf stars) and
$\sim61$\,K ($\sigma_T=62$\,K, $N=18$ giant stars).
}   
{Our temperature scale is slightly hotter than that of AAM96 and RM05
for metal-rich dwarf stars but cooler than that of RM05 for
metal-poor dwarfs. We have performed an fully self-consistent IRFM 
in the 2MASS photometric system. For those who wish to
use 2MASS photometry and colour-temperature calibrations to derive
effective temperatures, especially for metal-poor stars, we recommend
our calibrations over others available in the literature.
In our implementation we avoid the transformation of the
2MASS $JHK_{\rm s}$ magnitudes to a different photometric system and
thus fully exploit the excellent internal consistency of the 2MASS
photometric system.}   


\maketitle

\section{Introduction}
\label{introduction}

The effective temperature is a function of the bolometric flux and the
angular diameter according to the equation 

\begin{equation}
T_{\rm eff}=(\frac{4}{\sigma})^{1/4}\theta^{-1/2}F_{\rm bol}^{1/4}
\label{def_teff}
\end{equation}

\noindent
where $\sigma$ is the Stefan-Boltzmann constant,
$\theta$ the angular diameter, and $F_{\rm bol}$ the bolometric
flux measured on the surface of the Earth. However, \emph{direct}
measure of angular diameters is restricted to relatively few
stars, especially for dwarf stars.
\citet[][interferometry]{ker04,ker08} and 
\citet[][transit observations]{bro01} have directly measured the
angular diameters of bright stars. Only recently, \citet{bai08} have 
used the CHARA interferometric array to provide measurements of
angular diameters of $\sim28$ dwarf and subgiant stars, although all
of them have metallicities [Fe/H]~$>-0.5$. 

A \emph{semi-direct} method of temperature determination is one that
makes use of Eq.~\ref{def_teff} but relies on model atmospheres,
rather than on a direct measure of the angular diameter.
The infra red flux method 
\citep[hereafter IRFM;][and references therein]{bla90} is especially 
adequate for determining the effective temperature of F, G and K stars.
The  IRFM was first introduced by
\citet{bla77} who proposed simultaneously determining
the effective temperature and the angular diameter of a star. The
basic idea is to use the monochromatic flux in the infrared since it is
mainly dependent on the angular diameter but is approximately
dependent only on the first power of \teffo, whereas the
integrated flux strongly depends on the temperature (proportional
to \tefo).

Popular \emph{indirect} methods for deriving effective temperatures
are the excitation equilibrium of  
\ion{Fe}{i} lines \citep[e.g.][]{san04,san05} and on fitting 
Balmer lines \citep[e.g.][]{Fuhr93,Fuhr94,bar02}. 
Temperatures based on \ion{Fe}{i} excitation equilibrium 
depend on the model assumptions, such as non-LTE effects,
especially in metal-poor stars \citep[see][]{tai99,shc01}.
Recently, \citet{bar07} has also raised concerns
about possible non-LTE effects on the wings of Balmer lines.
Both excitation equilibria \citep{asplund05} and Balmer lines
\citep[][in~prep.]{ludwig08} are also sensitive to granulation
effects. This makes such methods strongly model-dependent, 
which is an undesirable feature. 
However, temperatures derived from Balmer lines and \ion{Fe}{i}
excitation equilibria have the considerable advantage of being
reddening independent.

One of the motivations of this work is to investigate the
trend of Li abundances towards low metallicities ([Fe/H]~$ < -2.5$), 
using our own implementation of the IRFM. \citep{bon07} investigated
the {\em Spite plateau} at the lowest metallicities (down to
[Fe/H]=--3.3) and found marginal evidence that there could be an
increased scatter or even a sharp drop in the Li abundance at these
low metallicities. Determination of the baryonic density from the
fluctuations in the cosmic microwave background (CMB) by the WMAP
satellite \citep{spe03,spe07} implies a primordial Li abundance, which
is at least a factor of 3--4 larger than observed on the {\em Spite
plateau}, creating a conflict with the traditional interpretation of
the plateau\citep{sas82a,sas82b}. This discrepancy would be even
greater if the drop in the Li abundance versus metallicity were to be
confirmed \citep[see][in~prep.]{sbor08}.

One decade ago, \citet{bon97} determined Li abundances using the
IRFM temperatures of \citet{aam96a}. They investigated the different
Li abundance trends found with different temperature scales.
In particular, the temperatures of \citet{rya96}, which are based on
the IRFM implementation of \citet{magain}, are cooler than the IRFM
temperatures of \citet{aam96a}, provided 
$T_{\rm Alonso}-T_{\rm Ryan}\sim +10$\,K at [Fe/H]~$\sim -1.7$ and
$\sim +130$\,K at [Fe/H]~$\sim -3.3$. 
From this, \citet{bon97} concluded that the presence or absence of
trends in lithium abundance with \teff is strongly dependent on the
temperature scale adopted. 

\citet{mar04} applied their own IRFM implementation
\citep{ram05a} to deriving the effective temperature and Li abundances
for a sample of stars similar to that of \citet{rya96}. They find
individual temperature differences of up to 400--500\,K 
for the some stars with metallicity below $-3.0$ dex. 
More recently, \citet{bon07} have compared the temperatures
obtained from $H\alpha$ profiles to other temperature indicators,
among them those from the IRFM-based colour--temperature calibrations
of \citep{ram05a} and \citet{aam96b}.
When a reddening based on the \citet{sch98} maps is adopted, from the
\teffo:(V-K) calibration of \citet{aam96b}, the mean
difference $T_{(V-K)_{\rm AAM96}}-T_{\rm H\alpha}$ is only 8\,K with 
a standard deviation of 100\,K. However, if we use the \teffo:$(V-K)$
calibration of \citet{ram05b}, this mean difference is
$T_{(V-K)_{\rm RM05}}-T_{\rm H\alpha}$ of 265\,K, with a standard
deviation of 122\,K. 

\citet{ram05a} add as calibrators a small sample of
metal-poor stars mainly from \citet{chr04} and \citet{cay04}, 
and a larger sample of metal-rich stars from \citet{san04} to the
original sample of \citet{aam96a}, and computed new \teffo--colour
calibrations.  
Since the majority of calibrators shared by the two
samples, this large difference ($\sim250$\,K) 
between the calibrations of \citet{ram05a} and \citet{aam96b} at low
metallicity is {\em a priori} unexpected. 
One could argue that the models used by the two groups
are not exactly the same, however they must be very similar
(ATLAS 9 models with the same ODFs and microturbulent velocity).
Since the IRFM is only weakly dependent on 
the models adopted, as shown by the results of \citet{cas06},
who used both ATLAS and MARCS models, it seems unlikely that
this difference is rooted in the different models.  
We suggest instead that this is because \citet{ram05a} use 
2MASS\footnote{The 2MASS catalogue can be accessed at
http://www.ipac.caltech.edu/2mass/.} $JHK_{\rm s}$ magnitudes for 
the low metallicity calibrators; such magnitudes were then transformed
into the TCS system to merge them with the homogeneous
set of TCS photometry of \citet{aam96a}. 
The errors in the transformation between the 2MASS and the 
TCS systems are then added to the photometric error and may have
undesired effects on the final calibration.
We have no way to prove that this is indeed the case; however, 
to circumvent such problems in this paper, we propose a new
implementation of the IRFM, including the stars from \citet{aam96a,
aam99a} and \citet{ram05a}, but using the 2MASS photometry for all
calibrators rather than a mixture of 2MASS and TCS.
The 2MASS magnitudes are probably not as accurate
as the careful TCS photometry of \citet{aam96a}, 
but the internal consistency of the 2MASS 
photometry is very high, about 1-2\% \citep{cutri}. 

\begin{figure}
\centering
\includegraphics[width=8.5cm]{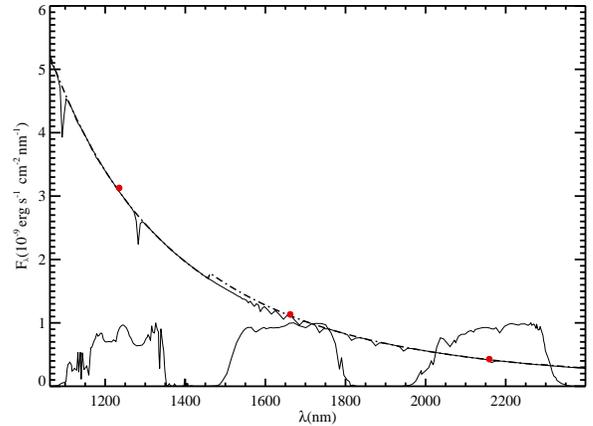}
\caption{Spectral energy distribution of the ATLAS model of Vega 
(\teff$= 9550$\,K, \logg$ = 3.95$, [Fe/H] $= -0.5$ and $v_{\rm
micro}=2$\,\kmso) in the infrared. The solid and dashed-dotted lines
represent the intensity and continuum flux, respectively. The filled
circles are the monochromatic fluxes adopted by \citet{coh03}. We also
show the transmission functions of the $JHK_{\rm s}$ 2MASS filters.}  
\label{specvega}
\end{figure}

\section{Implementation of the IRFM\label{impirfm}}

The IRFM \citep{bla90} evaluates the quotient between the bolometric
flux, $F_{\rm bol}$, and the monochromatic flux at a chosen infrared
wavelength, $F(\lambda_{\rm IR})$, both measured at the surface of the
earth, as an indicator of the \teffo. This quotient is the so-called
observational $R-$factor, $R_{\rm obs}$. The theoretical counterpart
derived from models, $R_{\rm theo}$, is obtained as the quotient
between the integrated flux, $\sigma$\tefo, and the monochromatic
flux at $\lambda_{\rm IR}$, $F_{\rm mod}(\lambda_{\rm IR})$, at the
surface of the star. Thus the basic equation of the IRFM is

\begin{align}
R_{\rm obs}&=\frac{F_{\rm bol}}{F(\lambda_{\rm IR})}=\frac{\sigma
T_{\rm eff}^4}{F_{\rm mod}(\lambda_{\rm IR}{\rm ,}T_{\rm eff}{\rm
,[Fe/H]}, \log g)} \notag\\
& =  R_{\rm theo}(\lambda_{\rm IR}{\rm ,}T_{\rm eff}{\rm
,[Fe/H]}, \log g) 
\end{align}

\noindent
where the dependence of models on metallicity, surface gravity, and
$\lambda_{\rm IR}$ is explicitly taken into account. The monochromatic
fluxes are obtained by applying the relation
\begin{equation}
F(\lambda_{\rm IR})=q(\lambda_{\rm IR}{\rm ,}T_{\rm eff}{\rm
,[Fe/H]}, \log g)[F_{\rm cal}(\lambda_{\rm IR})10^{-0.4(m_\star-m_{\rm
cal})}] 
\end{equation}

\noindent
where $m_\star$ is the magnitude of the target star,
and $m_{\rm cal}$ and $F_{\rm cal}$ are, respectively, the 
magnitudes and the absolute monochromatic fluxes of the calibrator
star (see Table~\ref{at9} and Sect.~\ref{zp}). The $q-$factor,
usually $\sim 1$, is a dimensionless factor that corrects the effect of the
different curvature of the flux distribution across the filter
window \citep[see][for more details]{aam94,aam96a,aam99a}. We have
used the definition of \citet{aam96a} for the computations of the
$q-$factors (see Sect.~\ref{zp}).

By merging the previous two equations we can separate the
observational and model inputs as

\begin{align}
\frac{F_{\rm bol}}{F_{\rm cal}(\lambda_{\rm
IR})10^{-0.4(m_\star-m_{\rm cal})}}= \qquad\qquad\qquad\notag\\
q(\lambda_{\rm IR}{\rm ,}T_{\rm eff}{\rm
,[Fe/H]}, \log g)R_{\rm theo}(\lambda_{\rm IR}{\rm ,}T_{\rm eff}{\rm
,[Fe/H]}, \log g)
\label{eqbasic}
\end{align}

\noindent
The synthetic magnitudes, the $q-$ and $R-$ factors,
necessary for implementing of the IRFM 
were computed from the ATLAS theoretical fluxes 
of \citet{castelligrid}\footnote{http://wwwuser.oats.inaf.it/castelli/grids.html}
using the 2MASS $JHK_{\rm s}$ filters. 
We used the fluxes in the ranges $3500$\,K$\,<\,$\teff$<7500$\,K,
$0.0<$\,\logg$<5.0$, and $-4<$\,[Fe/H]\,$<+0.5$, and
for metal-poor models with [Fe/H]~$\le-0.5$, we used
the fluxes computed from the $\alpha-$enhanced models. 
We derived a new calibration of the bolometric flux in the 2MASS
photometric system (see Sect.~\ref{bolf}). 

\begin{figure*}
\centering
\includegraphics[width=\textwidth]{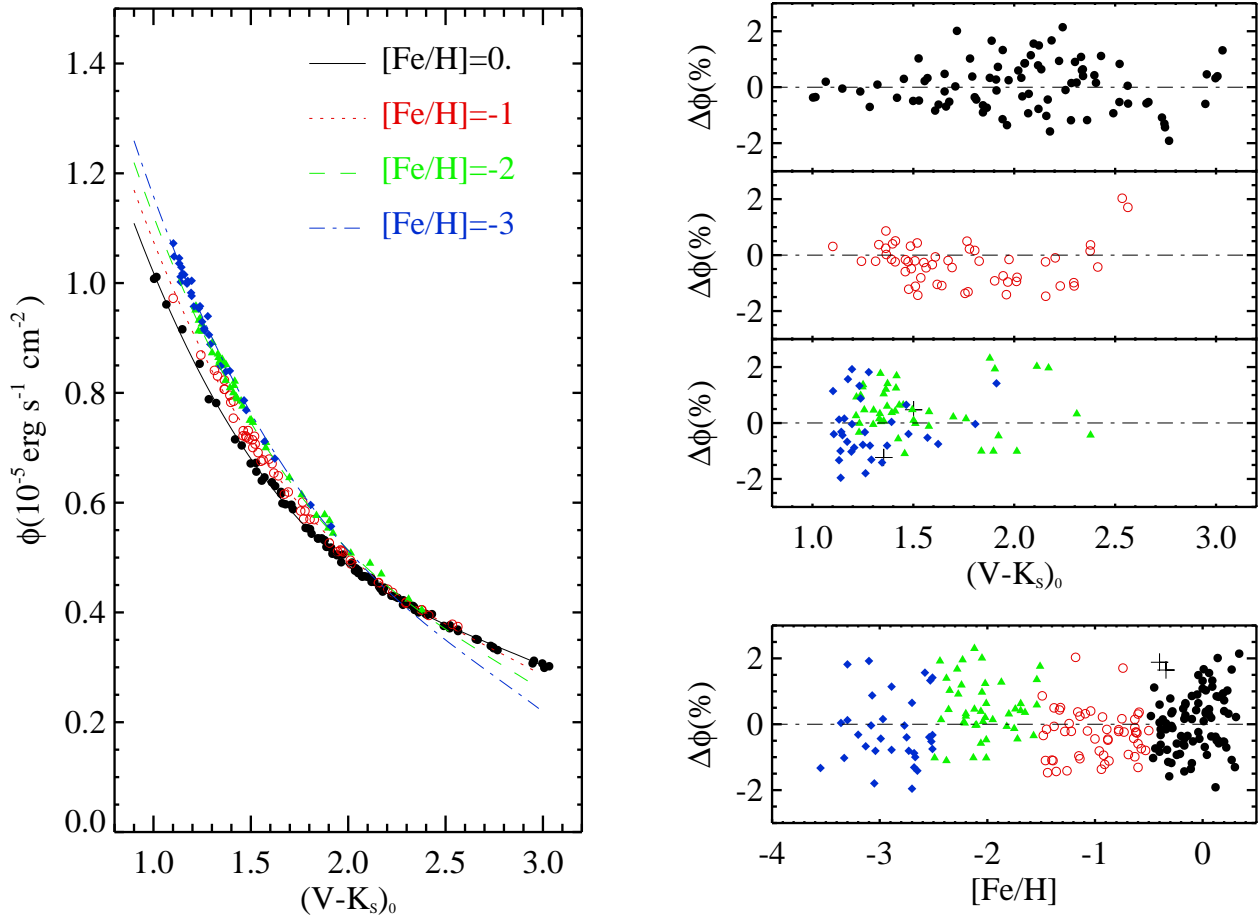}
\caption{{\em Left}: Empirical calibration $\phi$:$(V-K_{\rm s})$--[Fe/H] 
for dwarfs in the metallicity bins $-0.5 < [{\rm Fe}/{\rm H}] \le 0.5$ 
(filled circles), $-1.5 < [{\rm Fe}/{\rm H}] \le -0.5$ ({\em open circles}),
$-2.5 < [{\rm Fe}/{\rm H}] \le -1.5$ ({\em triangles}), and
$[{\rm Fe}/{\rm H}] \le -2.5$ ({\em diamonds}). The lines
correspond to our calibration for $[{\rm Fe}/{\rm H}]=0$ ({\em solid
line}), --1.0 ({\em dotted line}), --2.0 ({\em dashed line}), --3.0
({\em dotted-dashed line}). {\em Right}: Residuals of the fit
($\Delta \phi =(\phi_{\rm cal}-\phi_{\rm IRFM})/\phi_{\rm IRFM}$) as a function
of $(V-K_{\rm s})$ and [Fe/H].
}  
\label{figfbold}
\end{figure*}  

\section{Online data available at the CDS\label{online}}

\begin{figure*}
\centering
\includegraphics[width=\textwidth]{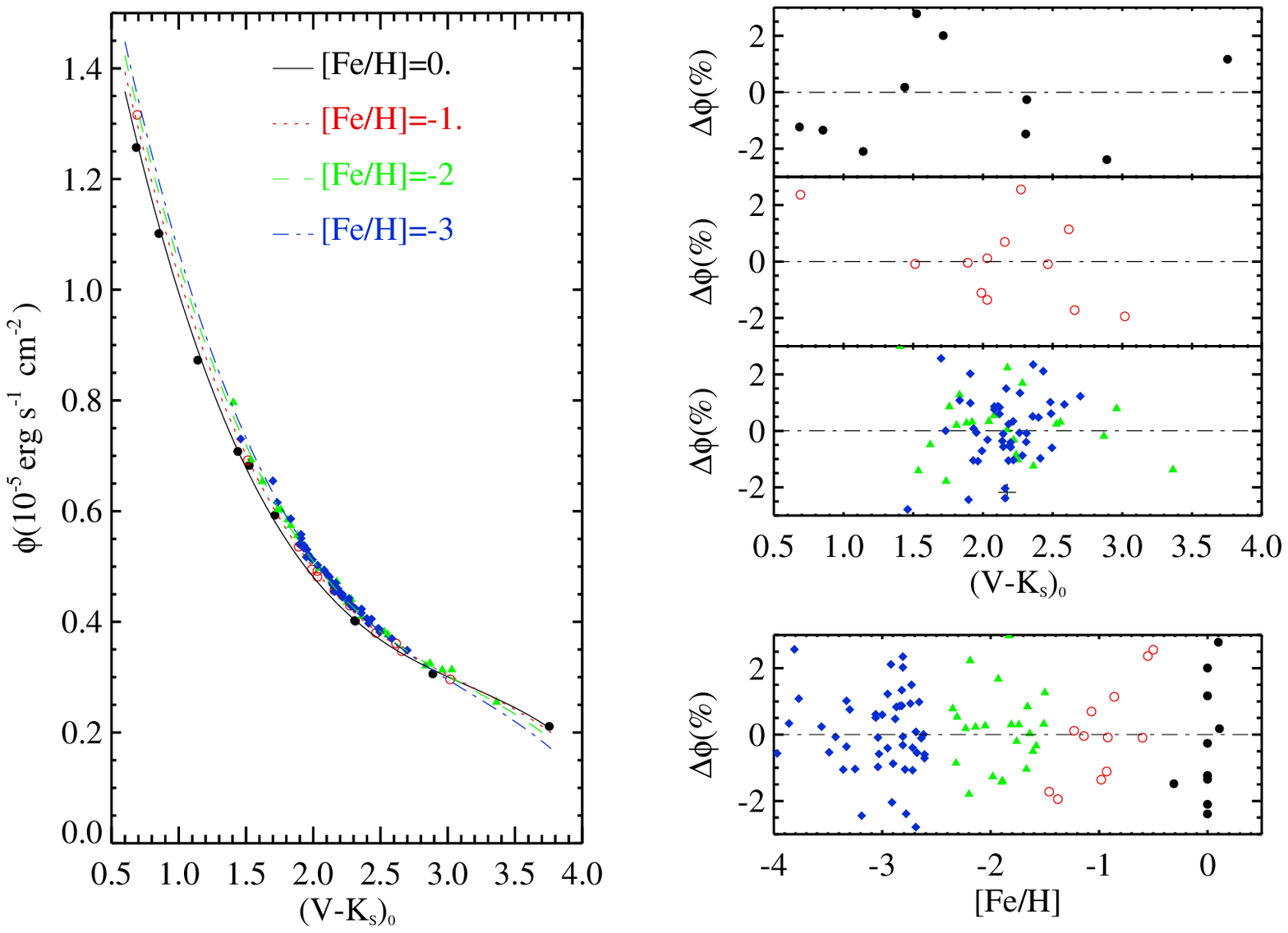}
\caption{The same as Fig.~\ref{figfbold}, but for giants.}  
\label{figfbolg}
\end{figure*}  

Several tables are available at the CDS
\footnote{http://cdsweb.u-strasbg.fr/}. We provide eight
tables containing the $q-$ and $R-$factors computed as indicated in
Sect.~\ref{impirfm} and~\ref{zp} for metallicities in the 
range [--4,0.5], temperatures in the range [3500,50000], and gravities 
in the range [0,5]. Within these tables, we also put the
theoretical colour V-K and magnitudes JHKs in the 2MASS system for
each atmospheric model. These theoretical colour and magnitudes, which
are not used in this work, weren normalised to
Vega assuming V=J=H=Ks=0. If the user wants to use a different
zero point for Vega, it is trivial to add it to
our theoretical magnitudes. In addition, eight
tables containing the $A-$factors and $BX-B$ coefficients 
for the same set of models are also available at the CDS,
needed for the bolometric flux calibration (see Sect.~\ref{bolf}).
At the CDS, we also provide two tables, with 555 dwarf
stars and with 264 giant stars, containing the photometric
data and reddenings used in this paper, stellar parameters and metallicity, bolometric fluxes
and IRFM temperatures for all the stars in our samples (see
Sect.~\ref{secphot} and~\ref{secmet}).  

\section{Sample, observational data, and stellar parameters}

\subsection{Photometric data\label{secphot}}

Our sample includes almost all the stars in \citet{aam96a,aam99a}
with available photometric data in the final release of the 2MASS
catalogue \citep{skr06} and with photometric accuracy $\lesssim
0.3$\,mag, for temperature determinations. We adopted this rather high
tolerance because giant stars of \citet{aam99a} are relatively bright and
usually the 2MASS photometric accuracy is very low for these stars.
However, to improve the precision of the bolometric flux
and \teffo:colour--[Fe/H] calibrations significantly, we decided to
further lower the accuracy limit down to $\lesssim 0.1$\,mag (see
Sect.~\ref{bolf} and~\ref{colourcal}). Therefore, stars with
2MASS photometric errors~$>0.1$ mag were only used for the purpose of
deriving effective temperatures and they are provided as online data
at the CDS. 

We adopted the same $UBV(RI)_{\rm C}$ photometric data as used by
\citet{ram05a,ram05b}, which were kindly provided by
Ram{\'\i}rez \& Mel\'endez (private communication). These data 
were extracted from the General Catalogue of Photometric Data 
\citep[GCPD]{mer97}.
For those stars of the \citet{aam96a,aam99a} samples without $V$
data in the GCPD these magnitudes were obtained from Simbad\footnote{
http://simbad.u-strasbg.fr/simbad/.}, and were later used to derive
bolometric fluxes and IRFM temperatures.

\subsection{Metallicity and surface gravity\label{secmet}}

For our sample of stars from \citep{aam96a,aam99a}, we adopted
the surface gravities and metallicities provided by Ram{\'\i}rez \&
Mel\'endez (private communication) which mostly use the mean values of
those reported in \citet{cds01}. 
 
We completed our sample of dwarfs and subgiants with the metal-rich
and metal-poor stars already included in the sample of \citet{ram05a}. 
The metal-rich sample mostly contains planet-host stars and
the comparison sample from \citet{san04}, but we also added to our
sample the stars with [Fe/H]~$>-2$ from \citet{cas06}. We completed
the sample with extremely metal-poor dwarfs from
\citet{bon07}, \citet{chr04}, and \citet{bar05}. For these stars, we
adopted the same surface gravity and metallicity as published in the
above papers.  

Our sample of giants contains the stars in \citet{aam99a},
plus the metal-poor stars from \citet{ram05a}. This 
includes stars from the ``First Stars'' project 
\citep{cay04,spi05}, and we adopted the surface
gravity and metallicity for these stars as provided in these papers. 

The errors on surface gravity and metallicity for all dwarf, subgiant,
and giant stars were assumed to be $\Delta \log g=0.5$ dex and
$\Delta$[Fe/H]~$=0.1$. The average systematic errors due to a 
different metallicity (by +0.1 dex) and a different surface gravity
(by +0.5 dex) are $13$\,K and $11$\,K, respectively, for dwarfs, and
$11$\,K and $28$\,K for giants. These errors were estimated by
quadratically adding the errors on effective temperature from each
band and calculating the average over all stars in both samples. 

\subsection{Reddening corrections\label{reddcorr}}

The extinction in each photometric band, $A_{i}$, as determined
using the relation $A_i=R_iE(B-V)$, where $R_i$ is given by the
coefficients provided in \citep{mcc04}. Reddening
corrections, $E(B-V)$, were adopted from Ram{\'\i}rez \&
Mel\'endez (private communication). For the metal-rich stars of
\citet{cas06} and the extremely metal-poor dwarfs of
\citet{bon07} and \citet{chr04}, reddening corrections were derived
from the maps of \citet{sch98}. The $E(B-V)$ from the maps is
appropriate for objects outside the dust layer, which is confined to
the Galactic disc. For objects which are within the dust layer
the map estimate should be corrected by a factor $[1-\exp(-|d\sin
b|/h)]$, where $d$ is the distance of the star,
$b$ its galactic latitude and $h$ the scale height of the
dust layer \citep[see, e.g.][]{calib}.
For this purpose we used the parallaxes 
provided by Simbad (which come mainly from the Hipparcos
catalogue \citealt{Perryman}) and assumed
a scale height of the dust layer of 125 \relax pc.
\citet{bon00} note that, when the maps of \citet{sch98}
provide reddenings larger than 0.1 mag, they
overestimate the reddening with respect to other
indicators, and proposed a simple formula
for correcting the reddening from the maps.
We make use of formula (1) of \citet{bon00}
to correct the reddenings derived
from the maps of \citet{sch98}.
 
\section{Photometric zero points and absolute flux 
calibrations for use with the IRFM\label{zp}}

Eq.~\ref{eqbasic} is what needs to be 
implemented practically to derive IRFM temperatures. 
The quantities on the lefthand side are observed quantities
while those on the righthand side are theoretical quantities.
One is immediately faced with a series of choices
\begin{enumerate}
\item the magnitude of the standard star ($m_{\rm cal}$)
\item the monochromatic flux of the standard star ($F_{\rm cal}$)
\item the zero point for $q$
\item the zero point for $R_{theo}$ 
\end{enumerate}

These choices are only apparently trivial. 
The 2MASS magnitudes have been carefully calibrated 
in absolute fluxes by \citet{coh03}; however, the standard
star to which the whole system is tied, Vega, has not been observed by
2MASS with sufficient accuracy due to its high brightness. 
A possible solution is to assume that the 2MASS magnitudes of Vega are
given by the zero points of Cohen et al., with changed sign, as in
\citet{cas06}. 
Another complication is the zero point of the theoretical
quantities. It is obvious from the definition of $q$ that its value
is 1 for the standard star; however, what are the
correct effective temperature, metallicity, and surface gravity 
of the standard star?
The Cohen et al. calibration relies on an ATLAS theoretical 
spectrum computed by R.L. Kurucz with the ``OLD'' opacity 
distribution functions assuming \teff = 9400, \logg = 3.9, a
metallicity of --0.5, and a microturbulent velocity of 0 \kms.  
Such a spectrum is not available in tabular form, we could
indeed recompute it, however using such a spectrum to zero
our theoretical quantities would mean using a spectrum that is
computed from a model inconsistent with the  
rest of the theoretical grid.
Furthermore, as we shall see in Sect.~\ref{bolf},
we will also need the absolute fluxes in other bands to
derive a calibration for the bolometric flux.
The natural choice is to use the corresponding theoretical
magnitudes of \citet{bcp98} transformed into the 2MASS system.
These magnitudes rely on the model for Vega, proposed by \citet{cak94},
consistent with the grid of \citet{castelligrid} that we are using.
A possible solution is to follow what was done by \citet{cas06},
who in fact used two different calibrations for optical and IR 
magnitudes.
Inspection of Eq.~\ref{eqbasic} suggests another solution:
use the same spectrum of the standard star to calibrate all bands.
In this way any error in the calibration will cancel out
when computing the flux ratio on the lefthand side of
Eq.~\ref{eqbasic}. However, to have a good absolute calibration,
one also needs accurate observed or derived 2MASS magnitudes of the
standard star Vega, which is quite difficult to obtain. 
We decided to adopt as 2MASS magnitudes of Vega those provided by
\citet[][see Sect.~\ref{bolf}]{mcc04}. 
This theoretical spectrum 
should also be used to define the zero point of $q$ and 
$R_{\rm theo}$ for the standard star, 
in order to have a fully self-consistent IRFM.

\begin{table}
\caption{Monochromatic Fluxes for Vega from the 
calibrated ATLAS 9 flux.\label{at9}}

\centering
\begin{tabular}{lrrr}

\hline\hline
Band&Wavelength  & Flux &  Mag Vega \\
    & nm &  $10^-9$&        \\
    &  & $\rm erg\,s^{-1}\,cm^{-2}\,nm^{-1}$\\ 
\hline\noalign{\smallskip}
J   & 1235 & 3.072 & 0.038\\
H   & 1662 & 1.113 & 0.040\\
K   & 2159 & 0.418 & 0.043\\
\hline
\end{tabular}
\end{table}

Throughout this work we adopt the theoretical flux of Vega of
\citet{cak94}
\footnote{http://wwwuser.oats.inaf.it/castelli/vega/fm05t9550g395k2odfnew.dat},
which has been calibrated to absolute flux, at Earth, using
the value recommended by 
\citet[$3.44\times 10^{-8}$ erg s$^{-1}$ cm$^{-2}$ nm$^{-1}$]{hay85}.
This spectrum is used to define the zero point of the $q$ factor
and the monochromatic fluxes, at the isophotal wavelengths of the
2MASS filters, listed in Table~\ref{at9} are used in our
implementation of Eq.~\ref{eqbasic}.
As noted by \citet{cas06}, such a calibrated
spectrum differs to the one used by Cohen et al. (1992) and
adopted by \citet{coh03} to define the absolute
flux calibration of the 2MASS magnitudes. In Fig.~\ref{specvega}
we display the calibrated spectrum of Vega in comparison with the
adopted monochromatic fluxes of \citet{coh03}. The difference is
small when comparing them with the continuum flux of our ATLAS 9
model of Vega at the same infrared wavelenghts.
We stress that, for the purpose of consistent
IRFM temperatures, we are not all that interested in having
accurate monochromatic fluxes, but instead accurate
ratios of bolometric fluxes to monochromatic fluxes. However, it
should be noted that the adopted observed magnitudes for Vega are
perhaps the source of uncertainty in a given temperature scale
based on the IRFM.

We adopted an error of 1 per cent on the monochromatic flux of each
band for the determination of effective temperatures. 
The average systematic errors due to this uncertainty are $44$\,K and
$46$\,K for dwarfs and giants, respectively. These errors were
estimated by adding  the errors quadratically on temperature from each
band and calculating the average over all stars in both samples.

\begin{table*}[!ht]
\caption[]{Coefficients and range of applicability of the
$\phi$:(V-m)--[Fe/H] calibrations.}  
\label{table.fbol}
\centering
\begin{tabular}{lccrrrrrrrrr}
\hline
\hline
\noalign{\smallskip}
Colour	& Colour range & [Fe/H] range & $a_0^{a}$ & $a_1$ & $a_2$ & $a_3$ & $a_4$ & $a_5$ & $a_6$ & N$^{b}$ & $\sigma_\phi(\%)^{c}$  \\       
\noalign{\smallskip}
\hline
\noalign{\smallskip}
\multicolumn{12}{c}{Dwarf stars} \\
\noalign{\smallskip}
\hline
\noalign{\smallskip}
$V-J$         & [0.8,2.4] & [--3.5,0.3] & 2.4945 & --2.2635 &   0.9615 & --0.1509 &  0.0657 & --0.1365 & --0.0074 & 219 & 0.9\\
$V-H$         & [0.9,3.0] & [--3.5,0.3] & 2.3681 & --1.9055 &   0.6415 & --0.0773 &  0.0418 & --0.1028 & --0.0053 & 216 & 0.9\\
$V-K_{\rm s}$ & [1.0,3.0] & [--3.5,0.3] & 2.3522 & --1.8817 &   0.6229 & --0.0745 &  0.0371 & --0.0990 & --0.0052 & 216 & 0.9\\
\noalign{\smallskip}
\hline
\noalign{\smallskip}
\multicolumn{12}{c}{Giant stars} \\
\noalign{\smallskip}
\hline
\noalign{\smallskip}
$V-J$         & [0.5,2.7]  & [--4.0,0.1] & 2.2282 & --1.7818 & 0.6809 & --0.0923  &  0.0302 &  --0.0696 & --0.0031 & 97 & 1.3\\
$V-H$         & [0.6,3.4]  & [--4.0,0.1] & 2.1522 & --1.5792 & 0.4821 & --0.0523  &  0.0182 &  0.0502 & --0.0019 & 91 & 1.3\\
$V-K_{\rm s}$ & [0.7,3.8]  & [--4.0,0.1] & 2.1304 & --1.5438 & 0.4562 & --0.0483  &  0.0132 &  0.0456 & --0.0026 & 95 & 1.4\\
\noalign{\smallskip}
\hline     
\end{tabular}
\begin{list}{}{}
\item[$^{a}$] The coefficients of the calibrations $a_i$ are given in units of 
10$^{-5}$ erg cm$^{-2}$ s$^{-1}$.
\item[$^{b}$] $N$ is the remaining number of stars after several
iterations (usually less than 20) of the 2.5$\sigma$ clipping.
\item[$^{c}$] $\sigma_\phi$, given in per cent, is the standard
deviation of the final calibrations.
\end{list}
\end{table*}

\section{Bolometric fluxes\label{bolf}}

One of the fundamental observational quantities for applying the IRFM is
the bolometric flux. The bolometric flux is not readily available for
any given star, but \citet{bap91} suggested that one could use a relation
of the type $F_{\rm bol} = 10^{-0.4m}\phi(X,{\rm [Fe/H]})$,
where $m$ is a suitable broad band colour and
$X$ a colour index. Such a calibration has
been derived by \citet{aam95} using the 
K magnitude and the $V-K$ colour (Johnson system);
\citet{cas06} have derived several similar calibrations for different
choices of $m$ and the colour index. 
In an initial attempt we tried to use the
\citet{aam95} calibration for this purpose,
which provided satisfactory results; 
however, the referees have correctly pointed out that, 
in doing so, we were forced to transform 
our ($V-K_{\rm s}$) colour into Johnson's system, 
thus losing the internal consistency
of the 2MASS system. Furthermore, we had to apply the
\citet{aam95} calibration outside its formal range of applicability,
for very metal-poor stars. The calibration of \citet{aam95}
only had two stars at [Fe/H]~$= -3.2$ and $-2.9$ and the rest with
[Fe/H]~$>-2.6$. A similar extrapolation problem would apply
if we had used any of the calibrations derived by \citet{cas06}, which
were derived for stars with [Fe/H]~$>-1.9$.
We therefore decided to derive a new calibration that makes use of
the $K_{\rm s}$ magnitude and the 2MASS-based ($V-K_{\rm s}$) colour
and covers the metallicity range appropriate to our sample of stars.

\begin{table}
\caption{Adopted absolute integrated fluxes and magnitudes 
for Vega.\label{flux_mag}}
\centering
\begin{tabular}{lrr}

\hline\hline
Band &  Flux & mag(Vega)\\
     &  $10^-5$&        \\
     & $\rm erg\,s^{-1}\,cm^{-2}$\\ 
\hline\noalign{\smallskip}
U    & 0.267 & 0.024 \\
B    & 0.607 & 0.028 \\
V    & 0.321 & 0.030  \\
R    & 0.341 & 0.037  \\
I    & 0.167 & 0.033  \\
J    & 0.050 & 0.038  \\
H    & 0.028 & 0.040  \\
K    & 0.011 & 0.043  \\
\hline
\end{tabular}
\end{table}

We adopt an approach similar to that of \citet{aam95}
and \citet{cas06}, with a slight difference.
The above authors use a set of effective wavelengths
and monochromatic fluxes for Vega in order to define
the integrated flux within each broad band from the 
photometry and the magnitudes of Vega. 
From the definition of magnitude follows

\begin{equation}
F_* = F_{\rm Vega}10^{-0.4(m-m_{\rm Vega})}
\label{bandflux}
\end{equation}

\noindent
where $m$ is any photometric band.
Provided then that the integrated flux of Vega
in any given band is known, the integrated
flux for the target star may be simply derived
from its measured magnitude and the magnitude of
Vega.
In Table~\ref{flux_mag} we provide
our adopted integrated magnitudes for Vega
for the bands we are interested in, and $JHK_{\rm s}$
refer to the 2MASS colours. 
These integrated magnitudes were derived
by integrating the filter response
functions of \citet{bes90} for the optical bands
and \citet{coh03} for the 2MASS bands, 
over the theoretical flux of Vega.
Consistently, the magnitudes for Vega were taken
from \citet{mcc04}, which gives the model magnitudes of \citet{bcp98} 
for the optical bands. This author
calculates the 2MASS magnitudes of Vega using the 
IR absolute monochromatic fluxes from \citet{coh03}, which are 
in fact quite similar to our adopted absolute monochromatic fluxes
from the calibrated model of Vega (see Fig.~\ref{specvega}).
Initially we were going to adopt the magnitudes of Vega equal to zero
in the 2MASS bands, but when we
derived the IRFM temperatures, our scale of temperatures was 
$\sim120$\,K hotter than that of \citet{aam96a}, which we think is 
the best implementation of the IRFM available in the literature due to
its internal consistency in the whole range of metallicities from --3.0
to 0.5. By adopting the magnitudes of Vega given by \citet{mcc04},
this difference is reduced to $\sim60$\,K, which we consider more
appropriate (see Sect.~\ref{alonso}).
In addition, the remarks on zero points of Sect.~\ref{zp} also apply
here. This choice guarantees that any error in the absolute
calibration of bolometric fluxes and monchromatic fluxes will cancel
out in Eq.~\ref{eqbasic}. 

\begin{figure*}
\centering
\includegraphics[width=\textwidth]{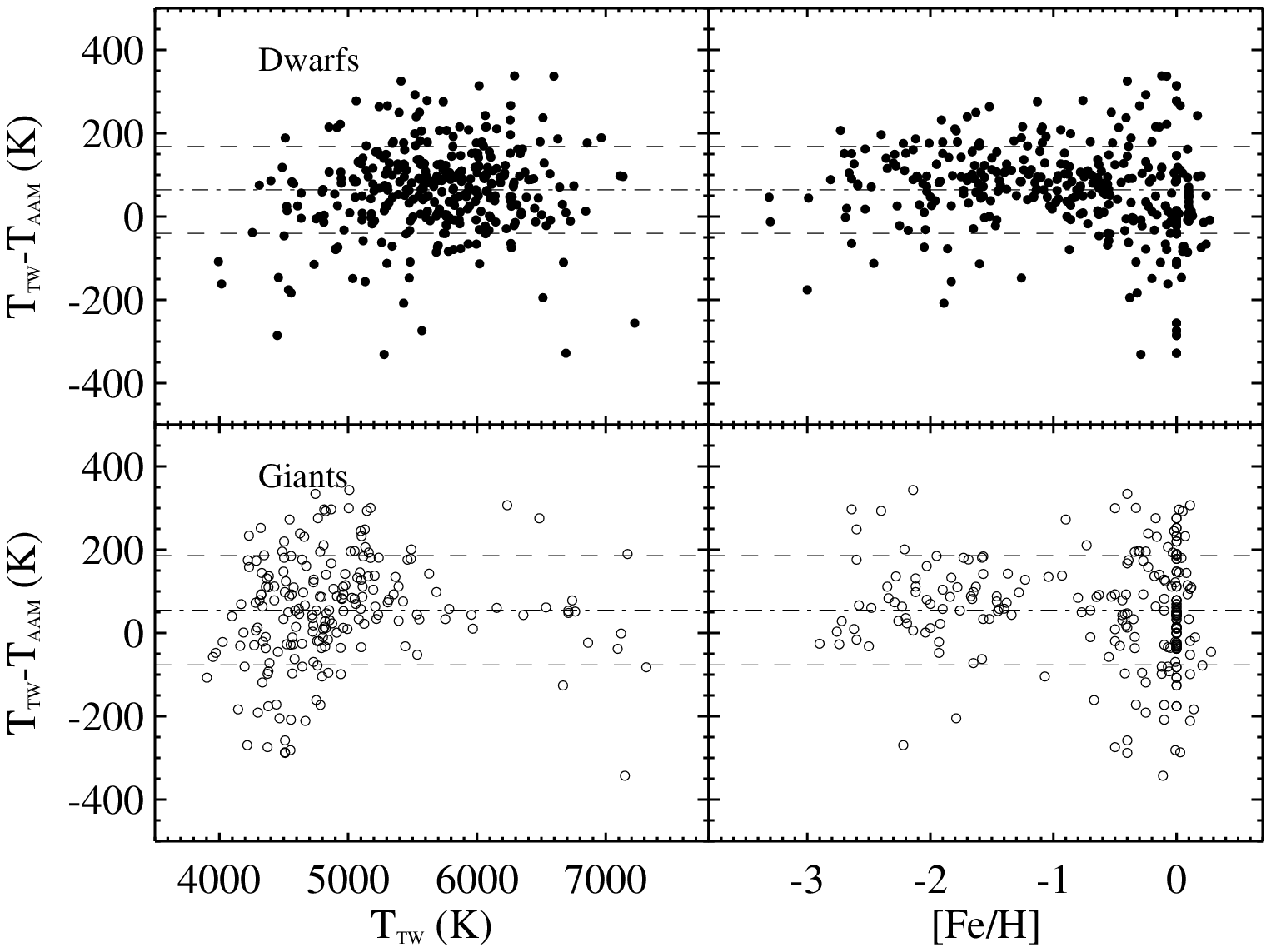}
\caption{Comparison of our temperature scale with that of
\citet{aam96a,aam99a}. The dashed-dotted line indicates the average
temperature difference and dashed lines the standard deviation,
$1\sigma$, from the average (see text).}    
\label{figAAM}
\end{figure*}  

\begin{figure*}
\centering
\includegraphics[width=\textwidth]{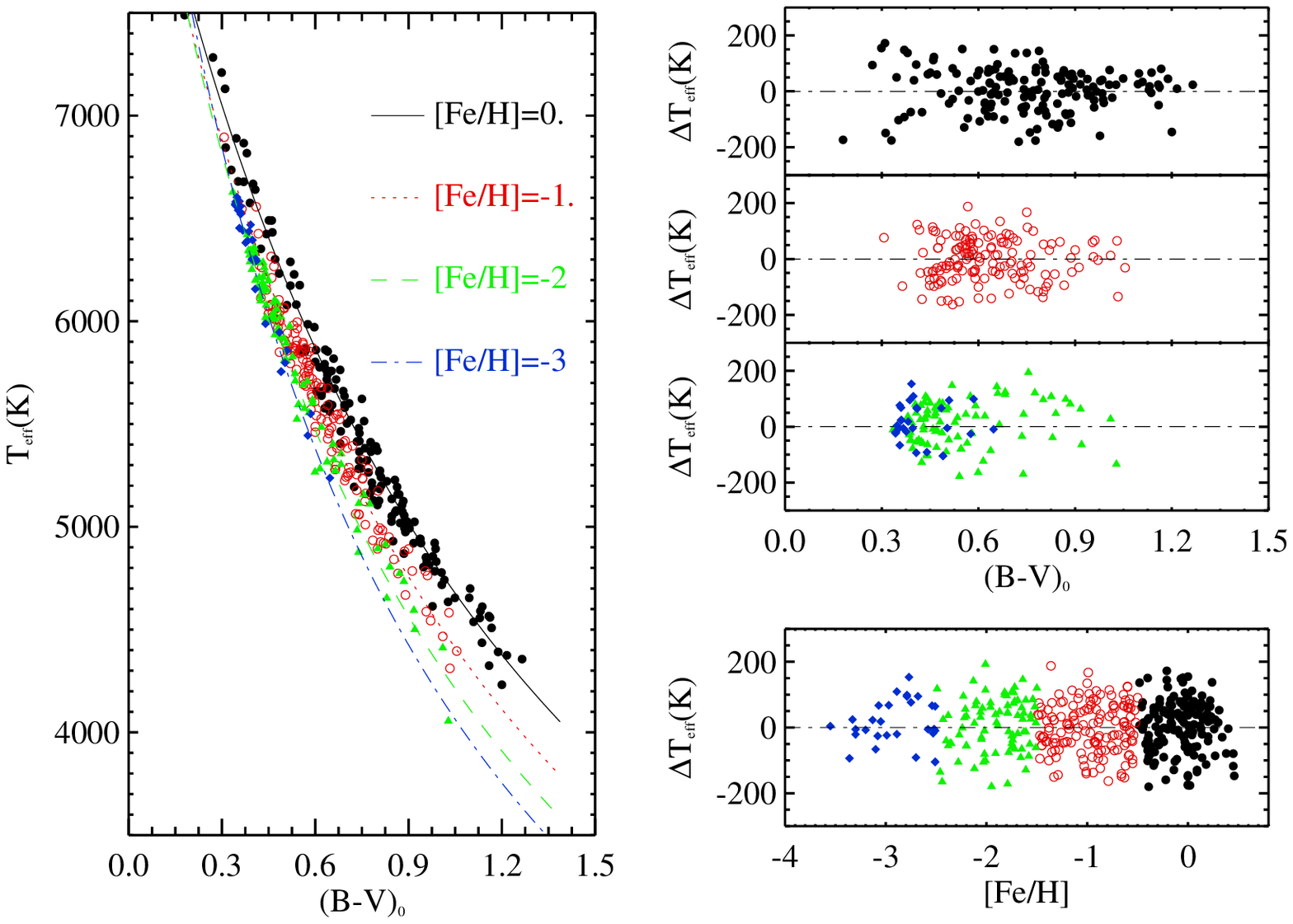}
\caption{{\em Left}: Empirical calibration \teffo:$(B-V)$--[Fe/H] for
dwarfs in the metallicity bins $-0.5 < [{\rm Fe}/{\rm H}] \le 0.5$ 
(filled circles), 
$-1.5 < [{\rm Fe}/{\rm H}] \le -0.5$ ({\em open circles}),
$-2.5 < [{\rm Fe}/{\rm H}] \le -1.5$ ({\em triangles}), and
$[{\rm Fe}/{\rm H}] \le -2.5$ ({\em diamonds}). The lines
correspond to our calibration for $[{\rm Fe}/{\rm H}]=0$ ({\em solid
line}), --1.0 ({\em dotted line}), --2.0 ({\em dashed line}), --3.0
({\em dotted-dashed line}). {\em Right}: Residuals of the fit
($\Delta T_{\rm eff}=T_{\rm eff}^{\rm IRFM}-T_{\rm eff}^{\rm cal}$) as a function
of $(B-V)$.
}  
\label{figBVd}
\end{figure*}  

From Eq.~\ref{bandflux} and the data in Table~\ref{flux_mag}
for any star for which photometry in several bands is available, one
may computed the total flux measured at Earth in the given bands. From
this value, the bolometric flux can be obtained by using model data.

In practice the stars we decided to use to derive the bolometric flux
calibrations fall into one of four groups:
\begin{enumerate}
\item
stars with full $UBV(RI)_{\rm C}JHK_{\rm s}$ data,
\item
stars with $UBVJHK_{\rm s}$ data,
\item
stars with $BVJHK_{\rm s}$ data,
\item
stars with $BV(RI)_{\rm C}JHK_{\rm s}$ data.
\end{enumerate}

The information is more complete for the stars of group 1) and 4);
however, we decided to include the also the stars of groups 2) and 3)
in our calibration effort, since this allows us to include a relevant
number of stars at extremely low metallicity. 

For each group one may compute
\begin{equation}
F_{obs} = \sum_i \int F(\lambda) T_i(\lambda) d\lambda
\end{equation}

\noindent
where the sum is extended to all the available bands
and $T_i(\lambda)$ is the response function of the
$i-th$ band, the integrals must be considered extended
from $0$ to infinity, formally,  although in practise the response
function of each filter vanishes outside a finite interval 
and numerically one stops integrating outside this interval.    
One can  then use the models to compute the correction

\begin{equation}
F_{\rm bol} = A F_{obs}
\label{afbol}
\end{equation}

\noindent
where $A$ is a function of \teff, \logg\ and [Fe/H].
Obviously a different $A$ has to be computed for
any given set of available bands. The $A$ factors 
for the different band combinations we have used are 
given in machine readable form at the CDS (see Sect.~\ref{online})
with different subscripts 1 to 4 corresponding to the different
band combinations.
This is again slightly different from what 
was done by \citet{aam95} or \citet{cas06}, 
who used the monochromatic fluxes at the effective wavelengths
of each band to approximate the spectral energy distribution
of the star and integrated this approximate energy distribution
over the whole interval. We only make use of integrated fluxes, which
are related to the observed magnitudes through Eq.~\ref{bandflux}
and of the fluxes and magnitudes of Vega given in
Table~\ref{flux_mag}. 

While the 2MASS magnitudes are provided by the catalogue, it is
customary for the optical bands to provide the $V$ magnitude and the  
colours $U-B$, $B-V$, $V-R$, $V-I$.
While for the bands $B$ to $I$ it is straightforward to
obtain the magnitude (e.g. $R = V-(V-R)$), some caution
must be exerted for the $U$ band, for which the atmospheric
extinction is strongly varying across the band and establishes the UV
cut-off. In fact, this band has proved to be the most difficult
to standardize. \citet{bes90} provides a response curve
$UX$ in which the atmospheric extinction is folded in 
and gives the curves $BX$ and $B$ for the $B$ band.
The former is to be used to compute the synthetic photometry
of the $(U-B)$ colour, while the latter is to be used
to compute the synthetic $(B-V)$ colour.
For the stars to be used in our bolometric flux calibration
we define the $U$ magnitude
\begin{equation}
U = (U-B) + V +(B-V) + (BX-B)
\end{equation}

\noindent
where $V$, $(U-B)$, and $(B-V)$ are the observed
magnitude and colours of the star, and $(BX-B)$
is derived from the theoretical models
with the requirement that it must be equal to zero for Vega. The
colours $(BX-B)$ are given in the online data at the CDS (see
Sect.~\ref{online}). 

The bolometric flux of each star was determined as in
\citet{aam95}. We first determined the fluxes of each band by applying
Eq.~\ref{bandflux} from the observed magnitudes of the star. Then
we derived the bolometric flux, $F_{\rm bol}$, using the
Eq.~\ref{afbol}. Thus the temperature \teff was then
determined using the IRFM that combines $F_{\rm bol}$ and the
monochromatic fluxes at IR wavelengths. This new value for the 
effective temperature may re-enter in Eq.~\ref{afbol} 
to derive a new
value for $F_{\rm bol}$, and so on. This iterative procedure converges
quickly towards a final $F_{\rm bol}$. In each iteration, the factors
$A$ and $BX-B$ were determined using a trilinear interpolation within
the grid for the corresponding \teffo, \loggo, and [Fe/H] of the star.
We considered the errors on the IRFM \teff due to uncertainties
on the adopted absolute calibration of the 2MASS photometric system,
and the errors on the magnitudes $JHK_{\rm s}$ \loggo, and [Fe/H].

We derived relations between bolometric fluxes and colours, 
taking also the effects of metallicity into account.
We adopted the same fitting formula as adopted by
\citet{cas06}

\begin{eqnarray}
\phi(X,{\rm [Fe/H]}) & = & a_0 + a_1 X + a_2 X^2 + a_3 X^3 +{}  \nonumber\\
 & & {}+a_4 X {\rm [Fe/H]} + a_5 {\rm [Fe/H]} + a_6 {\rm [Fe/H]}^2 
\label{fitb}
\end{eqnarray}

\noindent
where the $\phi$ is derived as $\phi(V-m,{\rm [Fe/H]})=
F_{\rm bol}/10^{-0.4m}$ ,~$X=V-m$ represents the $J$,$H$,$K_{\rm s}$
magnitudes, and $a_i$ $(i=0,...,6)$ are the coefficients of the fit. We
iterate the fitting procedure by discarding the points more than
2.5$\sigma$ from the mean fit. We also tried other fitting formulae
such as that of \citet{aam95}, but they led to similar results. This
has been extensively tested by \citet{cas06}, who also give fits using
optical bands. 

\begin{figure*}
\centering
\includegraphics[width=\textwidth]{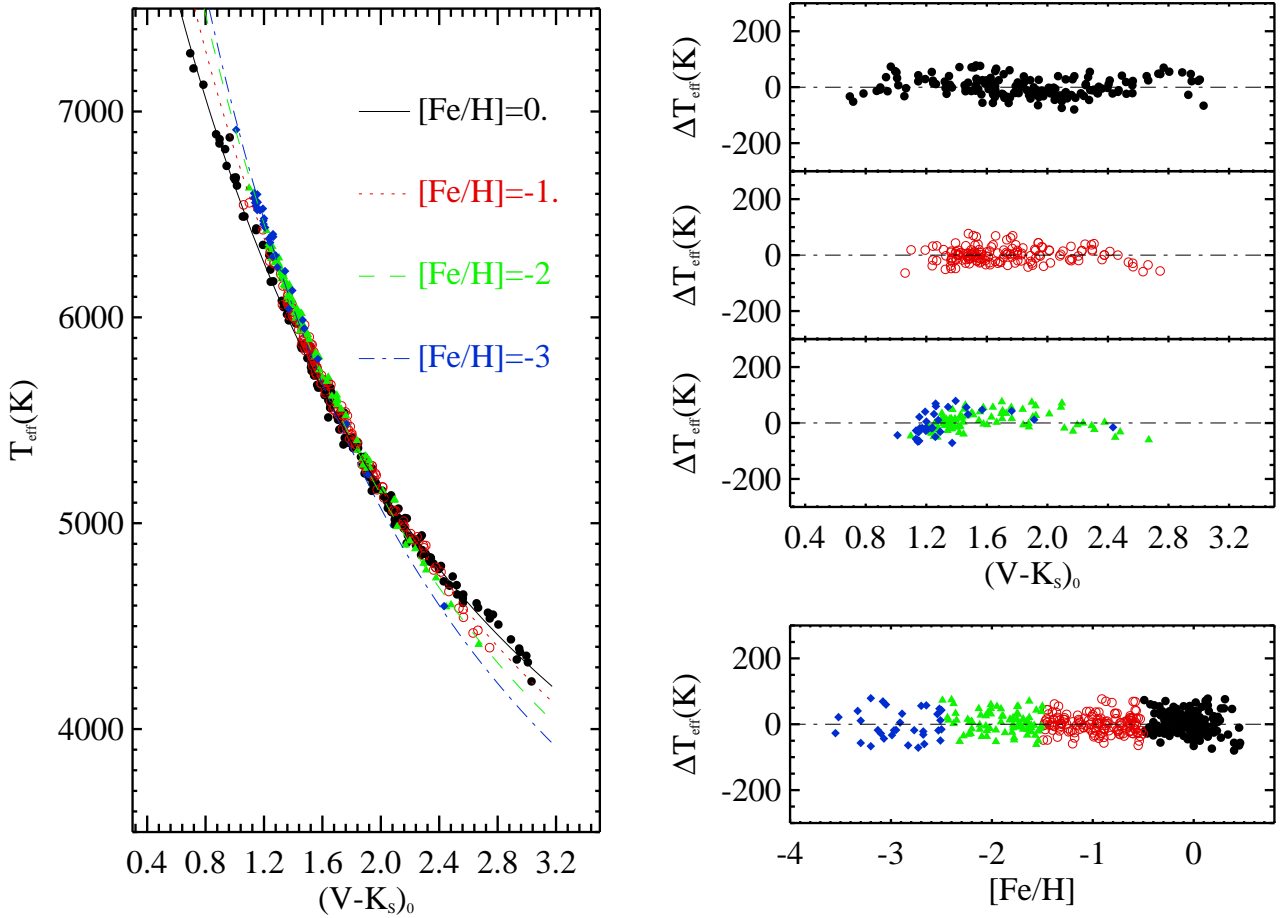}
\caption{The same as Fig.~\ref{figBVd}, but for $V-K_{\rm s}$ in
dwarfs.}  
\label{figVKd}
\end{figure*}  

For dwarf stars, we selected all the stars of group 1) from the sample
of \citet{aam96a} and \citet{ram05a} with uncertainties in the 
$JHK_{\rm s}$ magnitudes less than 0.1 at all metallicities. We
added the stars of group 4) from the sample of \citet{cas06} with the
same restrictions and we completed the sample with stars of
groups 2) and 3) with [Fe/H] $< -2$. We only added metal-poor stars to
give more weight to the metal-rich stars of groups 1) and 4);
otherwise, we would end up with including all dwarf stars and the fit
would be dominated by the greater number of stars of group 3).

For giant stars, we applied the same restrictions but the number of
stars with \citet{aam96a} and \citet{ram05a} of groups 1) and 4) was
very small (only 10 stars in group 1) and 26 in group 4) with
uncertainties in the $JHK_{\rm s}$ magnitudes less than 0.1), so we
decided to include all the stars of groups 2) and 3). 

In Figs.~\ref{figfbold} and~\ref{figfbolg} we
display the polynomial fits that represent the empirical
calibrations $\phi$ versus colours and metallicity. The coefficients
of these calibrations are given in Table~\ref{table.fbol}, together
with the remaining number of stars after the 2.5$\sigma$ clipping and
the r.m.s. of the fit, $\sigma_\phi$. These calibrations show
similar behaviours to those of \citet{aam95} and \citet{aam99a}, at
least in the metallicity range from $-3$ to 0.

\begin{table}
\caption[]{Comparison with other temperature scales.}  
\label{table.comp}
\centering
\begin{tabular}{lcrrr}
\hline
\hline
\noalign{\smallskip}
Sample	& [Fe/H] range & $\Delta T_{\rm eff}$ & $\sigma_{T_{\rm eff}}$ & N$^{a}$ \\       
\noalign{\smallskip}
\hline
\noalign{\smallskip}
\multicolumn{5}{c}{Dwarf stars} \\
\noalign{\smallskip}
\hline
\noalign{\smallskip}
\citet{aam96a} & [--3.5,+0.3] & +64 & 104 & 332 \\
\citet{aam96a} & [--3.5,-2.5] & +61 & 91  & 18  \\
\citet{aam96a} & [--0.5,+0.3] & +32 & 130 & 122 \\
\citet{ram05a} & [--4.0,+0.3] & +33 & 98 & 84 \\
\citet{ram05a} & [--4.0,-2.5] & --87 & 194  & 12  \\
\citet{ram05a} & [--0.5,+0.3] & +45 & 91 & 69 \\
\citet{cas06}  & [--1.9,+0.4] & --12 & 56 & 101 \\
\citet{cas06}$^b$  & [--1.9,+0.4] & --41 & 50 & 101 \\
\citet{san04}  & [--0.7,+0.5] & +11 & 120 & 133 \\
\citet{san04}$^b$  & [--0.7,+0.5] & --13 & 129 & 133 \\
\citet{bon07}  & [--3.6,-2.4] & +165 & 79 & 16 \\
\citet{bar02}  & [--2.5,+0.1] & +77 & 133 & 23 \\
\citet{bar02}  & [--0.5,-0.1] & +51 & 129 & 16 \\
\citet{chr04}  & [--3.1,-1.6] & +177 & 80 & 8 \\
\citet{bai08}$^c$  & [--0.4,0.5] & --32 & 163 & 22 \\
\noalign{\smallskip}
\noalign{\smallskip}
\hline
\noalign{\smallskip}
\multicolumn{5}{c}{Giant stars} \\
\noalign{\smallskip}
\hline
\noalign{\smallskip}
\citet{aam99a} & [--3.0,+0.5] & +54 & 131 & 202 \\
\citet{aam99a} & [--3.0,-2.5] & +76 & 120 & 10 \\
\citet{aam99a} & [--0.5,+0.5] & +43 & 144 & 116 \\
\citet{ram05a} & [--4.0,+0.3] & +63 & 57 & 25 \\
\citet{ram05a} & [--4.0,-2.5] & +61 & 62 & 18  \\
\citet{ram05a} & +0.2$^d$ & +116 & -- & 1  \\
\citet{cay04}  & [--4.0,-2.0] & +115 & 76 & 34 \\
\citet{chr04}  & [--3.4,-2.6] & +128 & 71 & 22 \\
\citet{bai08}$^c$  & 0$^e$ & --67 & 139 & 6 \\
\noalign{\smallskip}
\hline     
\end{tabular}
\begin{list}{}{}
\item[$^{a}$] The number of stars.
\item[$^{b}$] If we consider all reddening corrections equal to zero.
\item[$^{c}$] $\Delta T_{\rm eff}=T_{\rm
eff}^{\rm IRFM}-T_{\rm eff}^{\rm dir}$, where $T_{\rm eff}^{\rm dir}$
is a direct determination of \teff using the angular diameter
$\theta$. 
\item[$^{d}$] One metal-rich giant star.
\item[$^{e}$] Did not find any metallicity determination so decided to
adopt [Fe/H]~$=0$. 
\end{list}
\end{table}

\section{IRFM temperatures and angular diameters}

To determine effective temperatures we need to apply 
Eq.~\ref{eqbasic}. The bolometric fluxes are estimated using
the empirical calibration 
$F_{\rm bol,cal} = 10^{-0.4K_{\rm s}}\phi(V-K_{\rm s},{\rm [Fe/H]})$ given
in Table~\ref{table.fbol} and the 2MASS $K_{\rm s}$ and
Johnson $V$ magnitudes. The $q-$ and $R-$factors are determined from an
initial guess of the temperature of the star, \teffo$_0$, by 
trilinear interpolation in the grid, using the surface gravity 
and metallicity of the star.
Then, we determine a new value for the temperature by comparing the
theoretical bolometric flux, $F_{\rm bol,theo}$, derived from the
previous determination of \teff and the bolometric flux, 
$F_{\rm bol,cal}$, using the expression: 
\teffo$_{\rm ,new}=$\teffo$_{\rm ,old}[F_{\rm bol,cal}/F_{\rm
Bol,theo}]^{1/4}$. We again derive the $q-$ and $R-$factors for
\teffo$_{\rm ,new}$ and repeat this process iteratively until
$|$\teffo$_{\rm ,new}-$\teffo$_{\rm ,old}| \le 0.1$\,K.

The final temperature of the star is determined as the average of the
three temperatures extracted from each of 2MASS filters weighted with the
inverse of their individual errors \citep[see][]{aam96a}. The error 
on the weighted mean is computed as 
\begin{math}
\Delta T_{\rm eff}=N/\sum \left(\Delta T_i\right)^{-1}  
\end{math}
where $\Delta T_i$ are the errors of the temperatures from 
the individual filters ($i=J$,$H$,$K_{\rm s}$) and $N=3$ is the number
of available temperatures. 
These errors $\Delta T_i$ 
account for the photometric errors of the observed $JHK_{\rm s}$ and
$V$ magnitudes, the error on the adopted absolute calibration for the
2MASS photometric system, and the uncertainties on surface gravity and
metallicity. To estimate $\Delta T_i$, we just add all
the individual errors of the $i$ band quadratically. 

The angular diameters have been calculated from
Eq.~\ref{def_teff} with the derived IRFM temperatures and
bolometric fluxes. Their errors were estimated by propagating a mean
error of 1.3\% in the bolometric fluxes and the errors on \teffo.

\section{Comparison with other temperature scales.}

In this section, we compare our temperature scale with other
temperature determinations based on different implementations of the
IRFM \citep{aam96b,aam99a,ram05a,cas06}, on the excitation equilibrium
of \ion{Fe}{i} lines \citep{san04}, and on the fitting of Balmer line
profiles \citep{bar02,bon07}. In Table~\ref{table.comp} we gather the
mean differences between our temperatures and those of different
samples, $\Delta T_{\rm eff}$, together with the standard deviation
(scatter) around the mean, $\sigma_{T_{\rm eff}}$.

\subsection{Alonso et al. sample\label{alonso}}

The updated temperatures do not differ significantly from those of
\citet[][see Fig.~\ref{figAAM}]{aam96a,aam99a}. Our temperature scale is
hotter than that of \citet{aam96a,aam99a} for both dwarfs and giants.
We find an average
difference $\Delta T_{\rm eff}=+64$\,K with a $\sigma_{T_{\rm eff}}=104$
\,K ($N=332$ dwarfs) and $\Delta T_{\rm eff}=+54$\,K with a 
$\sigma_{T_{\rm eff}}=131$\,K ($N=202$ giants). This 
translates into a mean \teff difference of $\lesssim1\%$.
Although not negligible, such differences are within the error bars
of the current temperature determinations, although the scatter, 
$\sigma_{T_{\rm eff}}$, seems to be quite large.
The different bolometric flux calibration, photometric data and
absolute flux calibration might be responsible for this small
difference between the two temperature scales. 
\citet{cas06} checked that using the absolute calibration of
\citet{aam95} and if using the TCS filters, their calibration and that
of \citet{ram05a} agree within 20\,K. However, to do
this exercise they had to transform the 2MASS magnitudes into the TCS
system, so their conclusions may be affected by these
transformations.

Even if we select subsamples of
different mean metallicity, the  
differences remain very small (see Table~\ref{table.comp}).
In conclusion, in the whole metallicity range, the systematic
difference between our temperature scale and that of
\citet{aam96a,aam99a} in dwarfs and giants is positive but smaller
than $+65$\,K, which is in fact less that the average of the
individual uncertainties in our calibration ($\lesssim82$\,K for
dwarfs and $\lesssim76$\,K for giants).  

\begin{figure*}
\centering
\includegraphics[width=\textwidth]{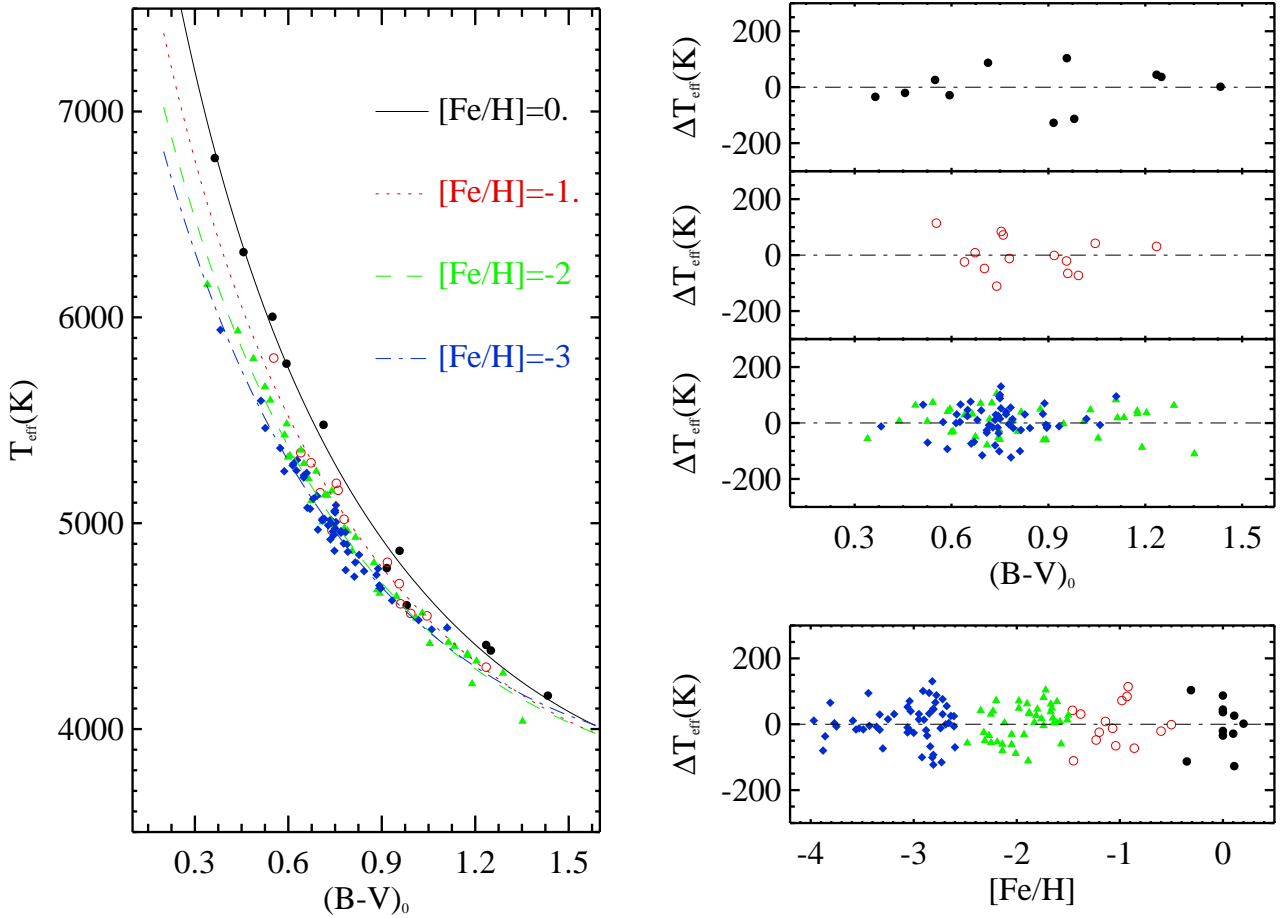}
\caption{The same as Fig.~\ref{figBVd}, but for giants.}  
\label{figBVg}
\end{figure*}  

\subsection{Ram{\'\i}rez \& Mel\'endez sample\label{sectram}}

\citet{ram05a} extend the sample of stars of \citet{aam96a,aam99a}
with metal-rich stars from \citet{san04} and very metal-poor stars
from \citet{chr04} and \citet{cay04}.
We determined effective temperatures for  the
calibrators of \citet{ram05a} using our implementation of the
IRFM. Our temperature scale is on average hotter than that of 
\citet{ram05a} by about $\Delta T_{\rm eff}=+33$\,K 
($\sigma_{T_{\rm eff}}=98$\,K, $N=84$ dwarfs) and 
$\Delta T_{\rm eff}=+63$\,K  
($\sigma_{T_{\rm eff}}=57$\,K, $N=25$ giants). This difference
might be partially related to the use of different absolute 
calibration as we stated in Sect.~\ref{alonso}.

Among giants, we find minor differences when we look at the most
metal-poor and metal-rich stars in the sample (see
Table~\ref{table.comp}).  
However, this behaviour changes when we inspect the dwarf stars.
While for metal-rich dwarfs we
find $\Delta T_{\rm eff}=+45$\,K  ($\sigma_{T_{\rm eff}}=76$\,K,
$N=69$ dwarfs with [Fe/H]~$> -0.5$), for metal-poor dwarfs we find
our temperature scale to be cooler:  
$\Delta T_{\rm eff}=-87$\,K  ($\sigma_{T_{\rm eff}}=194$\,K, 
$N=12$ dwarfs with [Fe/H]~$< -2.5$). 
We believe that this difference is mainly due to the photometric
transformations between the 2MASS and the TCS systems that
\citet{ram05a} need to perform in order to derive the IRFM
temperatures. 

\begin{figure*}
\centering
\includegraphics[width=\textwidth]{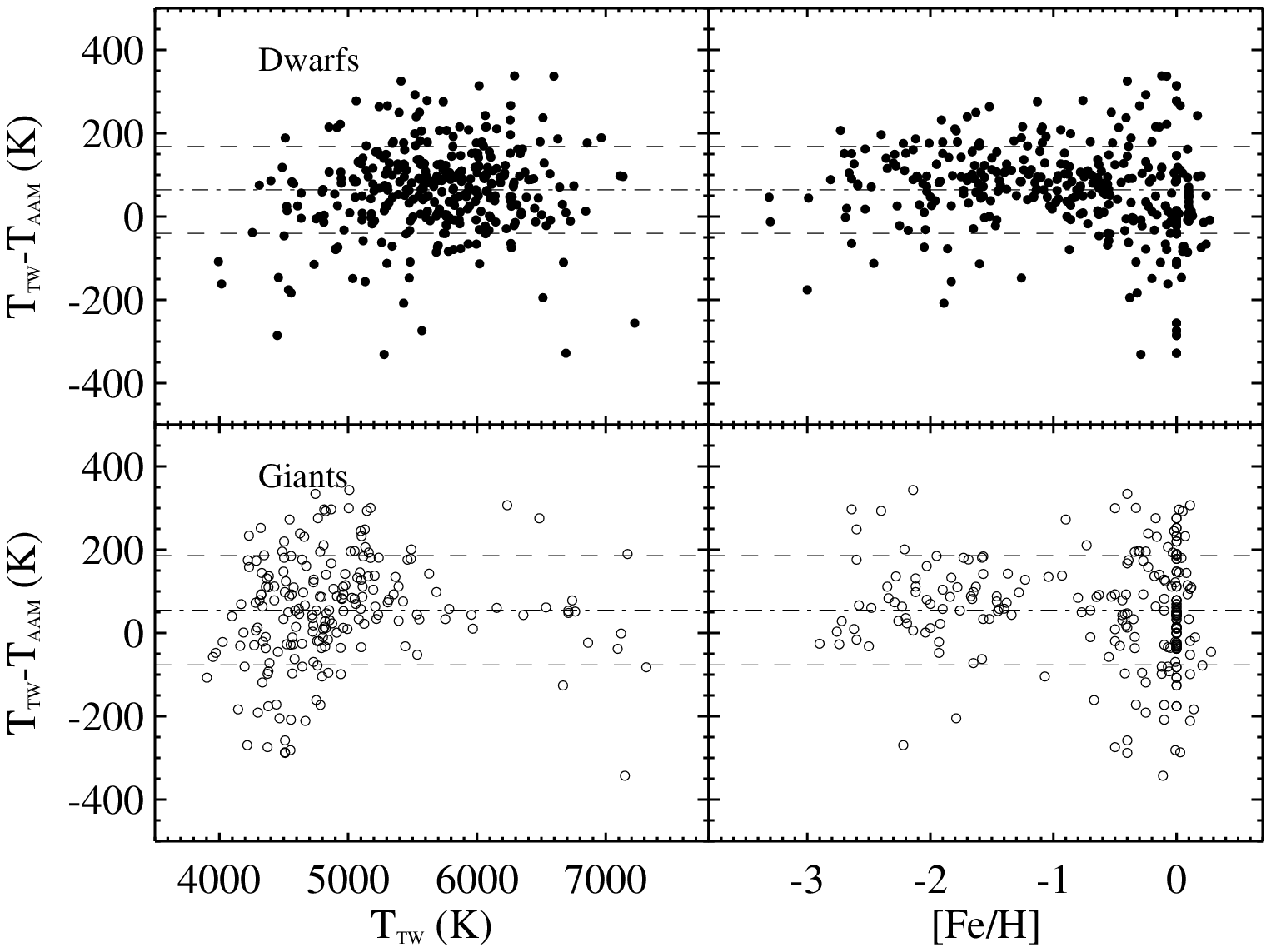}
\caption{The same as Fig.~\ref{figBVd}, but for $V-K_{\rm s}$ in
giants.}   
\label{figVKg}
\end{figure*}  

\subsection{Casagrande et al. sample}

\citet{cas06} propose a new IRFM using multiband
photometry. They derive empirical effective temperature and
bolometric flux calibration for G and K dwarfs stars in the
range $-1.87 <$[Fe/H]~$< 0.34$. They use $BV(RI)_{\rm C}$
Johnson-Cousins photometry and $JHK_{\rm s}$ 2MASS photometry. 
We applied our implementation to estimating the \teff of stars in
their sample and find our temperature scale only slightly cooler
by $\Delta T_{\rm eff}=-12$\,K ($\sigma_{T_{\rm eff}}=56$\,K,
$N=101$ dwarfs). For these stars, we estimated the reddening
corrections from the maps of dust of \citet{sch98}, 
corrected as described in Sect.~\ref{reddcorr}. In
Table~\ref{table.comp}, we also compare our temperature scale with
that of \citet{cas06} by arbitrarily adopting the reddening
corrections $E(B-V)=0$ for all the stars, under the assumption that
all these stars are nearby and should not show any reddening at
all. We find $\Delta T_{\rm eff}=-41$\,K ($\sigma_{T_{\rm eff}}=50$\,K,
$N=101$ dwarfs), i.e. temperatures 29\,K cooler on average. 
This systematic
difference is well within our error budget, so we decided to use these
reddening corrections to be consistent with other new stars included 
in the sample for which we need to estimate the reddening corrections
as the dwarf stars from \citet{bon07} (see Sect.~\ref{bonsec}).

\begin{table*}
\caption[]{Coefficients and range of applicability of the
\teffo:colour--[Fe/H] calibrations.}  
\label{table.colour}
\centering
\begin{tabular}{lccrrrrrrrr}
\hline
\hline
\noalign{\smallskip}
Colour	& Colour range & [Fe/H] range & $b_0$ & $b_1$ & $b_2$ & $b_3$ & $b_4$ & $b_5$ & N$^{a}$ & $\sigma_{T_{\rm eff}}({\rm K})^{b}$  \\       
\noalign{\smallskip}
\hline
\noalign{\smallskip}
\multicolumn{11}{c}{Dwarf stars} \\
\noalign{\smallskip}
\hline
\noalign{\smallskip}
$B-V$         & [0.2,1.3] & [--3.5,0.5] & 0.5725 & 0.4722 &   0.0086 & --0.0628 & --0.0038 & --0.0051 & 418 & 76\\
$V-R$         & [0.2,0.8] & [--3.1,0.3] & 0.4451 & 1.4561 & --0.6893 & --0.0944 & 0.0161 & --0.0038 & 164 & 45\\
$V-I$         & [0.5,1.4] & [--3.1,0.3] & 0.4025 & 0.8324 & --0.2041 & --0.0555 & 0.0410 & --0.0003 & 164 & 52\\
$V-J$         & [0.5,2.3] & [--3.5,0.5] & 0.4997 & 0.3504 & --0.0230 & --0.0295 & 0.0468 &   0.0037 & 430 & 36\\
$V-H$         & [0.6,2.8] & [--3.5,0.5] & 0.5341 & 0.2517 & --0.0100 & --0.0236 & 0.0523 &   0.0044 & 426 & 30\\
$V-K_{\rm s}$ & [0.7,3.0] & [--3.5,0.5] & 0.5201 & 0.2511 & --0.0118 & --0.0186 & 0.0408 &   0.0033 & 431 & 32\\
$J-K_{\rm s}$ & [0.1,0.8] & [--3.5,0.5] & 0.6524 & 0.5813 &   0.1225 & --0.0646 & 0.0370 &   0.0016 & 436 & 139\\
\noalign{\smallskip}
\hline
\noalign{\smallskip}
\multicolumn{11}{c}{Giant stars} \\
\noalign{\smallskip}
\hline
\noalign{\smallskip}
$B-V$         & [0.3,1.4]  & [--4.0,0.2] & 0.4967 & 0.7260 & --0.1563 &   0.0255  & --0.0585 & --0.0061 & 120 & 57\\
$V-R$         & [0.3,0.7]  & [--4.0,0.1] & 0.4530 & 1.4347 & --0.5883 & --0.0156  & --0.0096 & --0.0039 & 55 & 85\\
$V-J$         & [1.0,2.4]  & [--4.0,0.2] & 0.4629 & 0.4124 & --0.0417 & --0.0012  &   0.0094 &   0.0013 & 138 & 18\\
$V-H$         & [0.8,3.1]  & [--4.0,0.2] & 0.5321 & 0.2649 & --0.0146 & --0.0069  &   0.0211 &   0.0009 & 144 & 23\\
$V-K_{\rm s}$ & [1.1,3.4]  & [--4.0,0.2] & 0.5293 & 0.2489 & --0.0119 & --0.0042  &   0.0135 &   0.0010 & 145 & 23\\
$J-K_{\rm s}$ & [0.1,0.9]  & [--4.0,0.2] & 0.6517 & 0.6312 &   0.0168 & --0.0381  &   0.0256 &   0.0013 & 145 & 94\\
\noalign{\smallskip}
\hline     
\end{tabular}
\begin{list}{}{}
\item[$^{a}$] The remaining number of stars after several
iterations (usually less than 20) of the 2.5$\sigma$ clipping.
\item[$^{b}$] $\sigma_{T_{\rm eff}}$, given in K, is the standard deviation
of the final calibrations.
\end{list}
\end{table*}

\subsection{Santos et al. sample}

\citet{san04} have carried out a detailed spectroscopic analysis of
planet-host stars and a comparison sample of stars without known
planets. Their effective temperatures are based on the 
excitation equilibrium of the \ion{Fe}{i} lines. Our
temperature scale is only slightly hotter than that of \citet{san04}
with $\Delta T_{\rm eff}=+11$\,K ($\sigma_{T_{\rm eff}}=120$\,K,
$N=133$ dwarfs), although with a large scatter. As in the
previous section, we also derived the reddening corrections from the
\citet{sch98} dust maps, corrected as described
in Section \ref{reddcorr}. In Table~\ref{table.comp}, we also show 
the comparison with $E(B-V)=0$. 

\subsection{Cayrel et al. sample}

\citet{cay04} present UVES spectroscopic observations of very
metal-poor giant stars. They derived \teff using the
\teffo:colour--[Fe/H] calibrations of \citet{aam99b}. 
Our \teff scale is hotter by 
$\Delta T_{\rm eff}=+115$\,K ($\sigma_{T_{\rm
eff}}=76$\,K, $N=34$ giants with [Fe/H]~$<-2.5$); 
however, that part of this difference
stems from the different choice made for the reddening.
Here we adopted the reddening from the dust maps of \citet{sch98},
corrected as described in Sect.~\ref{reddcorr}, whereas
\citet{cay04} instead used the \citet{bah82} maps. The 
different choice in
reddening accounts for a difference of $\sim40-50$\,K, on average
\citep{cay04}. The remaining 75\,K reflect the difference
between our calibration and that of \citet{aam99b}.
Unsurprisingly, this is, essentially, the same as
what was found for giant stars with [Fe/H]~$<-2.5$ in
Sect.~\ref{alonso} (see Table~\ref{table.comp}).

\subsection{Bonifacio et al. sample\label{bonsec}}

\citet{bon07} present high quality spectroscopic data
of a sample of extremely metal-poor dwarf stars.
They derived the effective temperatures 
by fitting the wings of the $H\alpha$ line. We derived
the effective temperatures of these stars using the 2MASS $JHK_{\rm s}$
magnitudes and reddenings from \citet{sch98}, corrected
as described in Sect.~\ref{reddcorr}. Our effective
temperatures are significantly hotter than those derived from the
Balmer lines, $\Delta T_{\rm eff}=+165$\,K ($\sigma_{T_{\rm
eff}}=79$\,K, $N=16$ dwarfs with [Fe/H]~$<-2.5$). The
difference between the temperatures derived from $H\alpha$, and those
derived by using the colour $V-K$ in the calibrations of
\citet{ram05b} is roughly $265\pm122$\,K. This $\sim 100$\,K
difference may be partially explained by our comparison with the
temperature scale of \citet{ram05a} with $\Delta T_{\rm eff}=-87$\,K
(see Sect.~\ref{sectram}). This has an impact on the Li abundances
in extremely metal-poor stars down to [Fe/H]~$=-4$
\citep[see][in~prep.]{sbor08}, because the IRFM temperatures would
provide higher Li abundances at the lowest metallicities, whereas
$H\alpha$ temperatures seem to show a slowly decreasing trend in Li
towards lower metallicities. 

\citet{bon07} use the theory of \citet{barklem} 
to describe the self-broadening of Balmer lines.
For the same sample of stars, \citet{BIAU} instead use 
the \citet{AG65,AG66} theory and derived
effective temperatures which were on average 150\,K
hotter, thus in substantial agreement with our IRFM temperatures.

\subsection{Christlieb et al. sample}

\citet{chr04} present the Hamburg/ESO R-process Enhanced Star survey
(HERES) with the aim of searching for very metal-poor stars
([Fe/H]~$<-2.5$) with $r-$process elements enhanced. We selected those
stars with available $B-V$ and $V$ photometry in \citet{chr04} and 
took the stellar parameters from \citet{bar05}. The effective
temperatures were estimated by averaging the resulting \teff from the
different \teffo:colour--[Fe/H] calibrations of \citet{aam96b} and
\citet{aam99b}. They followed the prescription described by
\citet{siv04}. Our \teff scale is significantly hotter by 
$\Delta T_{\rm eff}=+177$\,K ($\sigma_{T_{\rm
eff}}=80$\,K, $N=8$ dwarfs with $-3.1$~[Fe/H]~$<-1.6$) and $\Delta
T_{\rm eff}=+128$\,K ($\sigma_{T_{\rm eff}}=71$\,K, $N=22$ giants
with $-3.4$~[Fe/H]~$<-2.6$). This difference probably comes from the
different adopted reddenings and the difference between our
temperature scale and that of \citet{aam96a} and \citet{aam99a}.

\subsection{Barklem et al. sample}

Balmer-line profile fitting in principle allows a very precise
determination of stellar effective temperature for cool stars.
\citet{bar02} claim an accuracy of the temperature
determinations of $\sim 65$\,K for solar metallicity stars but for
[Fe/H]~$\sim-1$ of $\sim 80$ and [Fe/H]~$\sim-2$ of $\sim 100$\,K. 
Uncertainties in the theory of self-broadening, 
deviations from LTE and granulation effects add
to the systematic error budget of Balmer-line based
effective temperatures.  
We compared our temperature scale with that of \citet{bar02}.
Our temperatures are
hotter by $\Delta T_{\rm eff}=+77$\,K ($\sigma_{T_{\rm eff}}=133$\,K, 
$N=23$ dwarfs with [Fe/H]~$>-2.5$). However, for metal-rich dwarfs with
[Fe/H]~$>-0.5$, this difference drops to $\Delta T_{\rm eff}=+51$\,K 
($\sigma_{T_{\rm eff}}=129$\,K, $N=16$ dwarfs with [Fe/H]~$>-0.5$). The
average difference between the two temperature scales remains within
the uncertainties on the temperature determinations, although the
standard deviation is large.

\section{\teffo:colour--[Fe/H] calibrations\label{colourcal}}

We derived relations between \teff and colours, also taking
the effects of metallicity into account.
We adopted the same fitting formula that was adopted by
\citet{aam96b,aam99b}, \citet{ram05b}, and \citet{cas06}

\begin{equation}
\theta_{\rm eff}= b_0 + b_1 X + b_2 X^2 + b_3 X {\rm [Fe/H]} + b_4 {\rm [Fe/H]} + b_5
{\rm [Fe/H]}^2
\label{fitf}
\end{equation}

\noindent
where $\theta_{\rm eff}=5040/T_{\rm eff}$,~$X$ 
represents the colour, and $b_i$ $(i=0,...,5)$  are the coefficients of
the fit. We iterate the fitting procedure by discarding the points
more than 2.5$\sigma$ from the mean fit. All our calibrations were
adequately tested by removing some terms and/or adding higher order
terms in either $X$ and [Fe/H]. We verified that neither removing
terms nor introducing higher order terms improves the
accuracy of the fit significantly. Therefore we adopted
Eq.~\ref{fitf}.  

In Figs.~\ref{figBVd},~\ref{figVKd},~\ref{figBVg}, and~\ref{figVKg},
we display the polynomial fits which represent the empirical
calibrations \teff versus colours and metallicity. We discarded all
the stars with uncertainties in the $JHK_{\rm s}$ magnitudes greater
than 0.1. The coefficient of
these calibrations are given in Table~\ref{table.colour}, along
with the remaining number of stars 
after the 2.5$\sigma$ clipping and the r.m.s. of the fit,
$\sigma(T_{\rm eff})$. Normally, the number of iterations were fewer
than 20. Our polynomial fits of the colour $B-V$ usually have similar
r.m.s. than those provided by \citet{ram05b}. We should point out the
small number of giant stars with metallicities [Fe/H]~$>-1.5$, because
most of the giant stars in the sample of \citet{aam99a} are very
bright objects, hence with poor-quality 2MASS $JHK_{\rm s}$
magnitudes. For dwarf stars, our empirical calibrations of $V-R$ and
$V-I$ have smaller r.m.s. than those of \citet{ram05b}. For giant
stars, the calibration of $V-R$ shows a greater r.m.s. than in
\citet{ram05b}, probably due to the small number of stars in our sample.

On the other hand, our empirical calibrations of the colours $V-J$,
$V-H$, and $V-K_{\rm s}$ have a smaller r.m.s. than those presented 
by \citet{ram05b}. 
For giants, our fits are more accurate although our sample contains
slightly fewer giant stars than the sample of \citet{ram05b}. 

\section{Angular diameters}

\begin{figure}[!tl]
\centering
\includegraphics[width=8.5cm]{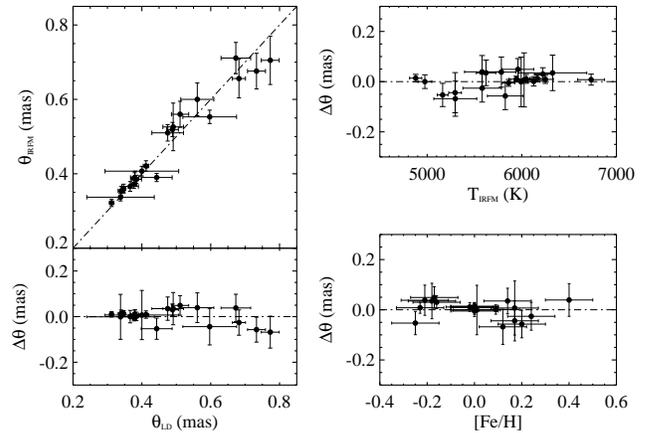}
\caption{Comparison between the angular diameters 
from \citet{bai08}, $\theta_{\rm LD}$, and diameters from the IRFM, 
$\theta_{\rm IRFM}$ for dwarf stars. Error bars are individual uncertaintes. 
The differences between $\theta_{\rm IRFM}$ and $\theta_{\rm LD}$,
$\Delta \theta$, are also shown containing the uncertainties of both
$\theta_{\rm IRFM}$ and $\theta_{\rm LD}$.} 
\label{figthetad}
\end{figure}  

\begin{figure}[!tr]
\centering
\includegraphics[width=8.5cm]{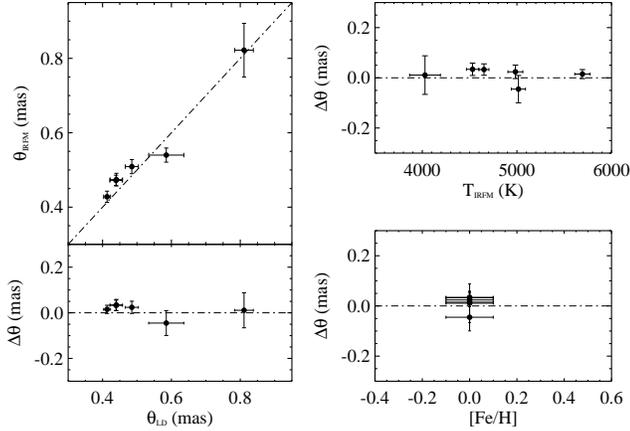}
\caption{Same as Fig.~\ref{figthetad}, but for giants.} 
\label{figthetag}
\end{figure}

The IRFM was developed to provide the \teff and $\theta$
simultaneously from observed and theoretical data. One fundamental
test to the IRFM is thus the comparison with measured angular
diameters. 
\citet{ram05a} compare their results with measured angular diameters
from \citet{rap02} and \citet{ker04} for giant and dwarf stars,
respectively. These stars are too bright for the 2MASS
catalogue, making the 2MASS $JHK_{\rm s}$ magnitudes very
uncertain. However, \citet{bai08} have recently presented new
measurements of angular diameters using the Center for High Angular
Resolution Astronomy (CHARA) Array, a six-element Y-shaped
interferometric array. We searched for the $JHK_{\rm s}$ magnitudes of
the stars reported in \citet{bai08} and the results are presented in
Figs.~\ref{figthetad} and~\ref{figthetag}. The stellar parameters were
adopted from \citet{bai08}, and the metallicities were extracted from
\citet{san04} and from \citet{cds01}. However, for same cases,
especially for giant stars, we did not find any available
metallicity determination, so we decided to adopt [Fe/H]~$=0$.  
The $V$ magnitudes were extracted from the GCPD \citep{mer97}, and
in those cases where no value was found, we took the $V$ magnitude as
given in the SIMBAD catalogue.
One can compare our IRFM angular diameters,
$\theta_{\rm IRFM}$, with the direct measurements, $\theta_{\rm LD}$.
For dwarf stars, the average difference, $\Delta \theta=\theta_{\rm
IRFM}-\theta_{\rm LD}$ is 0.002 with a standard deviation
$\sigma_\theta= 0.033$ ($N=22$ stars).
We can also derive a direct temperature, $T_{\rm eff}^{\rm dir}$, from
$\theta_{\rm LD}$ and the bolometric flux, determined from our
bolometric flux calibration, by using Eq.~\ref{def_teff}. The
previous comparison between angular diameters thus translates into a
temperature difference, $\Delta T_{\rm eff}=T_{\rm eff}^{\rm
IRFM}-T_{\rm eff}^{\rm dir}$, of $-32$ with $\sigma_{T_{\rm eff}}=
163$ ($N=22$ stars).  
For giants, the number of stars with relatively accurate $JHK_{\rm s}$
data is low. The sample of \citet{bai08} contains only six giant stars.
For these stars, we find $\Delta \theta=0.012$ with a $\sigma_\theta=
0.029$ which translates into $\Delta T_{\rm eff}=-67$ with a
$\sigma_{T_{\rm eff}}= 139$.  
These results are also given in Table~\ref{table.comp} in comparison
with other temperature determinations. 
Our new implementation of the IRFM provides 
good results when comparing with direct measurements of angular
diameters. 
The absence of any trend with metallicity in the 
residuals shown in Fig.~\ref{figthetad} over almost
1\,dex in metallicity is very encouraging. 
This suggests that the model atmospheres 
correctly model the variation of fluxes with metallicity.
Since the metal-rich range is the most difficult for
modelling the opacity, it is reasonable to 
expect that the models are also reliable 
at low metallicity.
In other words, we do not expect that our temperature
scale has spurious trends with metallicity due
to inadequate modelling of the stellar atmospheres.

\section{Summary}

We have made use of the IRFM to determine effective temperatures
of 555 dwarf and subgiant field stars and of 264 giant field stars.
Our implementation of the IRFM uses the 2MASS photometric system as a
reference system to perform all the calculations. We derived a
bolometric flux calibration down to metallicities of [Fe/H]~$=-3.5$ 
for dwarfs and $-4.0$ for giants, as a function of the 2MASS 
magnitude, $m$, and the colour, $V-m$. We computed theoretical
magnitudes by integrating the ATLAS models in the 2MASS $JHK_{\rm s}$
filters. 

Our temperature scale is hotter than that of \citet{aam96a,aam99a} by 
$\sim64$\,K  ($\sigma_T=104$\,K, $N=332$ dwarfs) and
$\sim54$\,K  ($\sigma_T=131$\,K, $N=202$ giants). Similar
results are found when comparing with \citet{ram05a}. However,
interestingly, for dwarfs stars with [Fe/H]~$<-2.5$, the 
comparison with the sample of \citet{ram05a} provides a difference of
$\sim -87$\,K  ($\sigma_T=194$\,K, $N=12$ dwarfs). 
We believe this difference is related to \citet{ram05a}
transforming the 2MASS $JHK_{\rm s}$ magnitudes to the TCS photometric
system to derive bolometric fluxes and effective temperatures
for their calibrators at the lowest metallicities, whereas we 
determine the bolometric fluxes and effective temperatures in the
2MASS photometric system. 

Our \teff are hotter than those estimated using $H\alpha$ line profiles
by $\Delta T_{\rm eff}=+77$\,K (\citealt{bar02},
$\sigma_{T_{\rm eff}}=133$\,K, $N=23$ dwarfs with [Fe/H]~$>-2.5$) and 
$\Delta T_{\rm eff}=+165$\,K (\citealt{bon07}, 
$\sigma_{T_{\rm eff}}=79$\,K, $N=16$ dwarfs with [Fe/H]~$<-2.5$). 
This result has implications for the Li abundances for very
metal-poor stars down to [Fe/H]~$=-4$. Higher temperatures provide
higher Li abundances; therefore, the drop of the Li abundances
towards lower metallicities will cancel out, and the Li abundances
would remain in a \emph{plateau}.

We derived \teff versus colour empirical calibrations,
which are compatible with those presented by 
\citet{aam96b,aam99b}, \cite{ram05b}, \citet{cas06}, 
within the quoted errors. 
For those who wish to use 2MASS photometry to estimate effective
temperatures for a wide range of metallicities, we recommend
our calibration, which were derived within
the 2MASS system, rather than the others, which are either
based on different systems or on hybrid systems.
Our calibrations exploit the excellent internal consistency
of the 2MASS photometry and should provide accurate
temperatures in a relative sense.
In an absolute sense, our calibrations are of the same
quality as the other calibrations.

A comparison of IRFM angular diameters with interferometric
measurements of angular diameters from \citet{bai08} provides 
good agreement for both dwarf and giant stars. This gives us
confidence that our new implementation of the IRFM is reasonable.

\begin{acknowledgements}

We are very grateful to Ivan Ram{\'\i}rez \& Jorge Mel\'endez for
kindly providing us with the photometric data that they collected in
2005 for dwarf and giant stars of Alonso et al. 
We acknowledge support from the EU contract
MEXT-CT-2004-014265 (CIFIST). This publication makes use of data
products from the Two Micron All Sky Survey, which is a joint project
of the University of Massachusetts and the Infrared Processing and
Analysis Center/California Institute of Technology, funded by the
National Aeronautics and Space Administration and the National Science
Foundation. This research has made use of the SIMBAD database,
operated at the CDS, Strasbourg, France.

\end{acknowledgements}

\end{document}